\documentclass[footinbib,twocolumn,amsmath,amstex,amssymb,mathfonts,superscriptaddress,prl]{revtex4-1}
\usepackage{graphicx}
\usepackage[usenames, dvipsnames]{color}
\usepackage{bm}
\usepackage{verbatim}
\usepackage[pdftex,colorlinks=red,bookmarks=false]{hyperref}
\usepackage{amsmath}
\usepackage{amssymb}
\usepackage{amsthm}
\usepackage{amsfonts}
\usepackage{soul}
\usepackage{float}
\usepackage{caption}
\usepackage{subfig}
\usepackage{bbm}

\begin{document}

\def\qv{\vec{q}}
\def\red{\textcolor{red}}
\def\blue{\textcolor{blue}}
\def\green{\textcolor{ForestGreen}}
\def\apricot{\textcolor{Apricot}}
\newcommand{\norm}[1]{\left\lVert#1\right\rVert}
\newcommand{\ad}[1]{\text{ad}_{S_{#1}(t)}}

\title{Anatomy of skin modes and topology in non-Hermitian systems}

\author{Ching Hua Lee}
\email{phylch@nus.edu.sg}
\email{calvin-lee@ihpc.a-star.edu.sg}
\affiliation{Institute of High Performance Computing, A*STAR, Singapore, 138632.}
\affiliation{Department of Physics, National University of Singapore, Singapore, 117542.}
\author{Ronny Thomale}
\affiliation{
 Institute for Theoretical Physics and Astrophysics, University of W\"urzburg, Am Hubland, D-97074 W\"urzburg, Germany
}


\date{\today}
\begin{abstract}
A non-Hermitian system can exhibit {\it extensive} sensitivity of its complex energy spectrum to the imposed boundary conditions, which is beyond any known phenomenon from Hermitian systems. In addition to topologically protected boundary modes, macroscopically many ``skin'' boundary modes may appear under open boundary conditions. 
We rigorously derive universal results for characterizing all avenues of boundary modes in non-Hermitian systems for arbitrary hopping range. For skin modes, we introduce how exact energies and decay lengths can be obtained by threading an imaginary flux. 
Furthermore, for 1D topological boundary modes, we derive a new generic criterion for their existence in non-Hermitian systems which, in contrast to previous formulations, does not require specific tailoring to the system at hand. Our approach is intimately based on the complex analytical properties of in-gap exceptional points, and gives a lower bound for the winding number related to the vorticity of the energy Riemann surface. It also reveals that the topologically nontrivial phase is partitioned into subregimes where the boundary mode's decay length depends differently on complex momenta roots. 
\end{abstract}
\maketitle 

The avenue of topological phases has reshaped our perspective on single-particle problems in condensed matter~\cite{haldane1988model,hasan2010colloquium,qi2011topological}. Unlike interacting many-body problems which are seldom exactly solvable, single-particle problems 
are often regarded as conveniently analytically tractable, with quantum and classical realizations accessible on equal formal footing~\cite{liang2013optical,chen2014experimental,nash2015topological,mousavi2015topologically,hu2015measurement,zhang2017entangled,lee2018topological,lohse2018exploring,helbig2019band,wang2019topologically}. 
This view, however, underestimates the richness and intricacies derived from the parameter and phase space structure of the physical system~\cite{Berry45}, as well as the added complexity implied by investigations of boundary terminations~\cite{dft}, external driving~\cite{doi:10.1080/00018732.2015.1055918}, and open systems beyond the realm of Hermiticity~\cite{ptnh}.

Non-Hermiticity from either inherent gain/loss or non-reciprocity is particularly interesting, exhibiting several exciting new phenomena. For instance, complex energy bands can develop branch cuts terminating at so-called exceptional points~\cite{berry2004physics,1751-8121-45-44-444016,lee2016anomalous, hu2017exceptional,achilleos2017non,shen2018topological,zhong2018winding,zhang2018dynamically} that can coalesce to form exceptional rings~\cite{zhen2015spawning,carlstrom2018exceptional,yang2019non}, and bulk modes can morph into boundary ``skin'' modes exhibiting an extensively large boundary density of states~\cite{yao2018edge,alvarez2018non}. Non-Hermiticity profoundly affects topological localization in fascinating, yet poorly understood ways. In a topologically non-trivial Hermitian system, a boundary can only introduce a sub-extensive number of in-gap protected modes. The bulk modes, being de-localized, remain largely undisturbed. 
In contrast, in a non-Hermitian system, the \emph{entire} spectrum of an arbitrary large system can be modified by introducing a boundary, ostensibly violating the bulk boundary correspondence (BBC)~\cite{lee2016anomalous,xiong2018does,alvarez2018non,yao2018edge,kunst2018biorthogonal,kawabata2018non}. 

As we shall elucidate, this seemingly counterintuitive sensitivity to boundary conditions is a consequence of the fundamental observation that non-reciprocal systems can be driven  into different regimes by local perturbations, each characterized by its distinct exceptional points and winding numbers. This is because non-reciprocity can localize all eigenmodes at the boundaries, including those which, for periodic boundary conditions, would have been assigned extended bulk modes. There are two types of non-Hermitian boundary eigenmodes: Extensive skin modes which are adiabatically connected to Hermitian bulk modes through complex analytic continuation, and sub-extensive topological boundary modes, which are, as we will show, protected by a universal non-Hermitian topological winding number criterion.

Recent attempts at characterizing these enigmatic non-Hermitian boundary modes have not always been conclusive. Even after generalizing the Berry curvature and Chern number to their biorthogonal non-Hermitian analogs~\cite{shen2018topological,yao2018non,gong2018topological, kawabata2018non}, difficulties remain in choosing the most appropriate and efficient quantities and contours for capturing phase transitions~\cite{gong2018topological}. While Refs.~\cite{yao2018edge} and~\cite{kunst2018biorthogonal} have identified jumps in the biorthogonal polarization as necessary conditions for topological phase transitions, their sufficiency remains unclear beyond the simplest models with nearest-neighbor hoppings. Since non-reciprocity fundamentally alters the non-Bloch energy spectrum, the eigenmodes of generic models with multiple non-reciprocal hopping ranges can only be understood through a systematic analysis of their \emph{complex} band structure. Quantitative predictions of the localization lengths and dispersions of skin modes are even more elusive, with existing results restricted to numerical evidence or fine-tuned models where boundary modes can be calculated exactly~\cite{kawabata2018non,alvarez2018non,kunst2018biorthogonal}. 
Thus, the key outstanding questions are: (i) What are not just necessary but also \emph{sufficient} conditions for the skin effect in non-Hermitian systems? (ii) How can one analytically characterize the energies, density of states, and localization lengths of skin modes? (iii) What is a universal criterion for topological boundary modes of 1D non-Hermitian systems that does not require specially tailored contours?
In this work, we answer these questions through a universal treatment of boundary modes in non-Hermitian systems.

\noindent{\it Complex flux for characterizing skin modes --}
Usually, open boundaries break translational invariance and preclude exact analytic characterization of the eigenmodes. For non-Hermitian skin modes, however, analytical results exist via a mode pumping argument~\cite{PhysRevB.23.5632,niu1990towards,PhysRevLett.77.570,soluyanov2011wannier, alexandradinata2011trace,lee2015free,hatsugai2016bulk}
with a \emph{complex} flux $\phi$. 
We propose to interpolate between periodic and open boundary conditions (PBCs and OBCs) by adiabatically reducing one of the boundary hoppings to zero via this complex flux. As a minimal model to illustrate the idea, consider a generic 1D chain with particle hopping of arbitrary range. In momentum space, it is represented by the Hamiltonian $H=\sum_{n=-N_L}^{N_R}\sum_{k}e^{ikn}T_n\eta^\dagger_{k}\eta^{\phantom{\dagger}}_{k}$, where $T_n$ is the hopping amplitude across $n$ unit cells, and $\eta^\dagger_{k}$ creates a particle with quasi-momentum $k$. (Note that $T_n$ becomes matrix-valued as soon as there are multiple states per unit cell.) When the hoppings are non-reciprocal, $T_n\neq T^T_{-n}$ ($H\neq H^T$ in real-space, see~\cite{malzard2015topologically}), hence allowing the left/right hopping ranges $N_L$ and $N_R$ to be not necessarily equal. 
To evolve from PBCs to OBCs, we first transform each hopping $T_n\rightarrow T_ne^{in\phi}$ through flux threading. Next, we perform a gauge transform $H\rightarrow V^{-1}HV$ with $V=\text{diag}(e^{-i\phi},e^{-2i\phi},...,e^{-il\phi})$, with $l$ being the system length, to remove the complex phase from all but the boundary hoppings, which are consequently multiplied by $e^{\mp i l \phi}\sim e^{\mp l\,\text{Im}\,\phi}$. Since skin modes are spatially localized, the divergent case $e^{l|\text{Im}\,\phi|}$ can always be ignored by choosing an appropriate sign~\cite{Suppmat} for $\phi$. We are hence left with boundary hoppings rescaled by $e^{-l|\text{Im}\,\phi|}$, which corresponds to perfect PBCs when $\phi=0$, and the OBC limit when $\phi\rightarrow \infty$. Implementing $\phi$ threading by minimal coupling $k\rightarrow k+i\kappa$, this implies that the \emph{translationally invariant} analytic continuation of the original Hamiltonian,
\begin{equation}
H_\kappa(k)=H(k+i\kappa),
\label{H_kappa}
\end{equation}
has the same spectrum as the skin modes due to boundary hoppings suppressed by $e^{-\kappa l}$ per unit hopping. Physically, \eqref{H_kappa} implies that 
all the original PBC bulk states \emph{must} morph into left boundary modes with localization lengths $\kappa^{-1}$ under $e^{-\kappa l}$ boundary hopping suppression. Furthermore, $H_\kappa(k)$ $\forall\,\kappa$ forms an equivalence class of Hamiltonians with identical OBC spectra~\cite{yao2018edge,jin2018bulk}. 

Our approach allows us to understand why superficially similar systems may still manifest markedly different non-Hermitian effects. We demonstrate this insight through the non-Hermitian Su-Schrieffer-Heeger (SSH) model~\cite{lee2016anomalous,yao2018edge,kunst2018biorthogonal,lieu2018topological,yin2018geometrical}: 
\begin{equation}
H_{\text{SSH}}^{\gamma_x,\gamma_y}(k)=\left(\frac1{2}+\cos k\right)\sigma_x+\sin k\sigma_y+i\gamma_x\sigma_x+i\gamma_y\sigma_y,
\label{gamma_SSH}
\end{equation}
\begin{figure}[H]
\begin{minipage}{\linewidth}
\includegraphics[width=\linewidth]{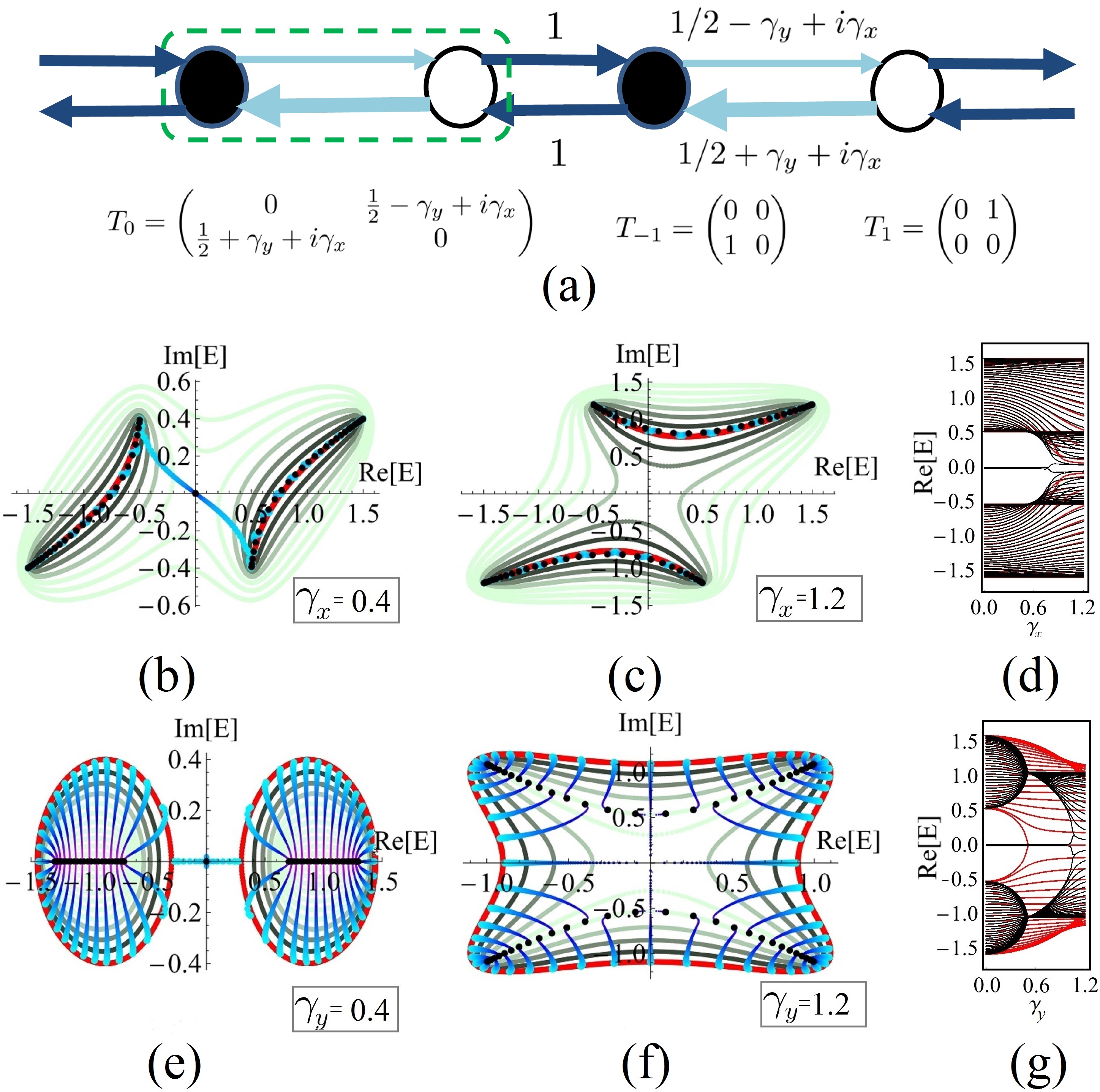}
\end{minipage}
\caption{a) Illustration of Eq.~\ref{gamma_SSH}, with unbalanced intra-unit cell couplings $T_0$ and balanced inter-unit cell couplings $T_{\pm 1}$. Spectra for $H_{\text{SSH}}^{\gamma_x,0}$ (b-d) and $H_{\text{SSH}}^{0,\gamma_y}$ (e-g) from Eq.~\ref{gamma_SSH}. (d,g): Both systems exhibit topological zero modes (black lines at $\text{Re}[E]=0$) at small $\gamma_x$ or $\gamma_y$, but  only $H_{\text{SSH}}^{\gamma_x,0}$ exhibits BBC. For $H_{\text{SSH}}^{0,\gamma_y}$, \emph{all} the OBC modes (black) differ from the PBC modes (red), not just for topological modes. (b,c,e,f): Anatomy of PBC and OBC spectra in a topologically nontrivial ($\gamma_{x,y}=0.4$) and trivial ($\gamma_{x,y}=1.2$) regime. Light blue-magenta tapering curves illustrate the evolution of PBC modes into OBC modes as $\text{Im}\,\phi$ increases from $0$ to $\infty$.  Pale background closed curves are contours of constant $\kappa=\text{Im}\,k$ with intervals of $0.1$. For $H_{\text{SSH}}^{\gamma_x,0}$, the PBC and OBC spectra coincide along open arcs and no evolution occurs, except towards the isolated topological zero mode in (b). For $H_{\text{SSH}}^{0,\gamma_y}$, the OBC skin modes (black) morph en-masse from the PBC modes (red). Bulk modes become localized skin modes as soon as they lie along $\kappa\neq 0$ contours. From Eq.~\ref{beta}, skin modes (black) of (f) satisfy  $\text{Re}[E^2]=\frac{5}{4}-\gamma_y^2$ and $4\gamma_y^2>1+(\text{Im}[E^2])^2$. }
\label{fig1}
\end{figure}
\noindent with $\sigma_x,\sigma_y$ denoting the Pauli matrices. The breaking of non-reciprocity relies exclusively on $\gamma_y$, which becomes transparent from the $T_{0,\pm 1}$ hopping matrix representation in Fig. ~\ref{fig1}a. As shown in Fig.~\ref{fig1}, $H_{\text{SSH}}^{\gamma_x,0}$ and $H_{\text{SSH}}^{0,\gamma_y}$ possess qualitatively different behavior as PBCs are morphed into OBCs via imaginary flux ($\kappa=\text{Im}\,\phi$) pumping. $H_{\text{SSH}}^{\gamma_x,0}$  (Fig.~\ref{fig1}b-d) respects the usual BBC, with its OBC (black) and PBC (red) spectra coinciding except for isolated topological boundary modes. For $H_{\text{SSH}}^{0,\gamma_y}$  (Fig.~\ref{fig1}e-g), however, almost the \emph{entire} spectrum collapses onto the OBC modes (black) when evolving towards OBCs, i.e. one finds the non-Hermitian skin effect. They all become boundary modes because only modes on the PBC loci (red in Fig.~\ref{fig1}c,d,f,g) have real Bloch momenta. In particular, PBC bulk modes tend to evolve into the interior of their PBC loci, and will not move (i.e. obey the usual BBC) only if already located along an open arc, as for $\gamma_y=0$. In the following, we shall explain and analytically characterize such behavior.

\noindent{\it OBC constraints and skin mode solutions --} One may be tempted to find skin modes simply by taking the $\kappa\rightarrow \infty$  OBC limit in Eq.~\ref{H_kappa}. This, however, would yield undetectable modes with vanishing decay lengths. To correctly find the skin modes of a Hamiltonian $H(z)$, where $z=e^{ik},\; k \in \mathbb{C}$, we construct an ansatz eigenmode $\psi(x)$ from the eigenenergy $E$ subset of the Hilbert space:
\begin{align}
\psi(x)&=\sum_{\mu}c_\mu\beta_\mu^x\varphi_\mu,\\
E\varphi_\mu&=H(\beta_\mu)\varphi_\mu,
\label{psi_x}
\end{align}
where $c_\mu$ denots complex coefficients and the set of $\beta_\mu$s consists of all the roots of the characteristic polynomial $\text{Det}[H(z)-E\,\mathbb{I}]=0$, $E$ regarded as a fixed parameter. The bulk Hamiltonian specifies that
\begin{equation}
H\psi(x)=\sum_{-N_L<n<N_R;\mu}c_\mu\beta_\mu^{x+n}T_n\varphi_\mu=E\psi(x),
\label{H_PBC}
\end{equation}
which is satisfied for any set of $c_\mu$s. Beyond that, the OBC places additional constraints stipulating that the mode $\psi(x)$ must vanish outside $x\in[0,l]$. Specifically, at every site at position $x_L$($x_R$) within a maximal distance of $N_L$($N_R$) sites from the left(right) edge, hoppings of range $n$, where $x_{L/R}\leq n\leq N_{L/R}$ (call them $n\in \Gamma_{L/R}$) that goes beyond the edge should be truncated from the Hamiltonian. Subtracting these boundary truncations for $0<x_L\leq N_L$ and $0<x_R\leq N_R$ from Eq.~\ref{H_PBC}, we obtain the following $N_L+N_R$ constraints for the left and right boundaries, respectively:
\begin{eqnarray}
\sum_{n\in \Gamma_L}z^{-n}T_{-n}\psi(x_L)&=&\sum_{n\in \Gamma_L; \,\mu} c_\mu\beta_\mu^{x_L-n}T_{-n}\varphi_\mu=0,\notag\\
\sum_{n\in \Gamma_R}z^{n}T_{n}\psi(l-x_R)&=&\sum_{n\in \Gamma_R;\,\mu} c_\mu\beta_\mu^{l-x_R+n}T_{n}\varphi_\mu=0.\qquad \label{OBC1}
\end{eqnarray}
They collectively determine the coefficients $c_\mu$~\cite{Suppmat}. In the thermodynamic limit of large $l$, only the largest $|\beta_\mu|$ term(s) survive in Eq.~\ref{OBC1}. Yet, there should generically be at least two equally large $|\beta_{\mu_{max}}|$ if $\varphi_{\mu_{max}}$ were to survive, 
since otherwise none of the other terms will be large enough to cancel the $c_{\text{max}}\beta_\mu^l$ term as $l\rightarrow \infty$. Exceptions occur when 
Eqs.~\ref{OBC1} is not full rank due to some fortuitous redundancies in the $T_{\pm n}\varphi_\mu$'s; such isolated cases will be revealed as ``topological'' modes later. Hence we conclude that for any non-topological bulk or skin boundary mode to exist, a \emph{necessary} condition is that:
\begin{equation}
\exists \,\,\,\mu\neq\nu\;\;\;\text{such that } \,\,|\beta_\mu|=|\beta_\nu|.
\label{beta}
\end{equation}
Eq.~\ref{beta} has previously appeared in Refs.~\cite{yao2018edge} and \cite{kunst2018biorthogonal} as the condition for an extended bulk state, where a topological phase transition leads to a biorthogonal polarization jump~\cite{kunst2018biorthogonal}. As evident above, however, our Eq.~\ref{beta} has a far broader scope: It is the condition for \emph{any} non-topological mode to exist under OBCs, be they bulk or skin modes. Note that Eqs.~\ref{OBC1} and hence \ref{beta} are valid regardless of Hermiticity: In particular, for Hermitian bulk modes, Eq.~\ref{beta} holds trivially since $|\beta_\mu|=1$ for all Bloch modes.
\begin{figure}[H]
\begin{minipage}{\linewidth}
\subfloat[]{\includegraphics[width=.48\linewidth]{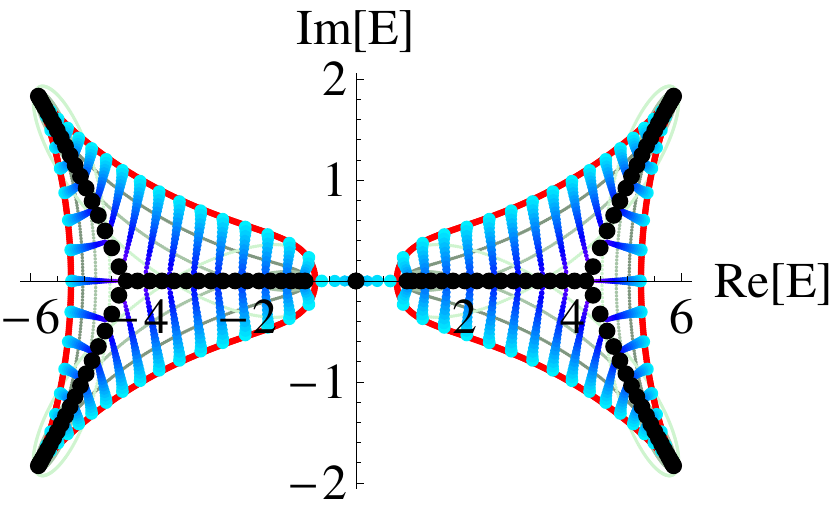}}
\subfloat[]{\includegraphics[width=.51\linewidth]{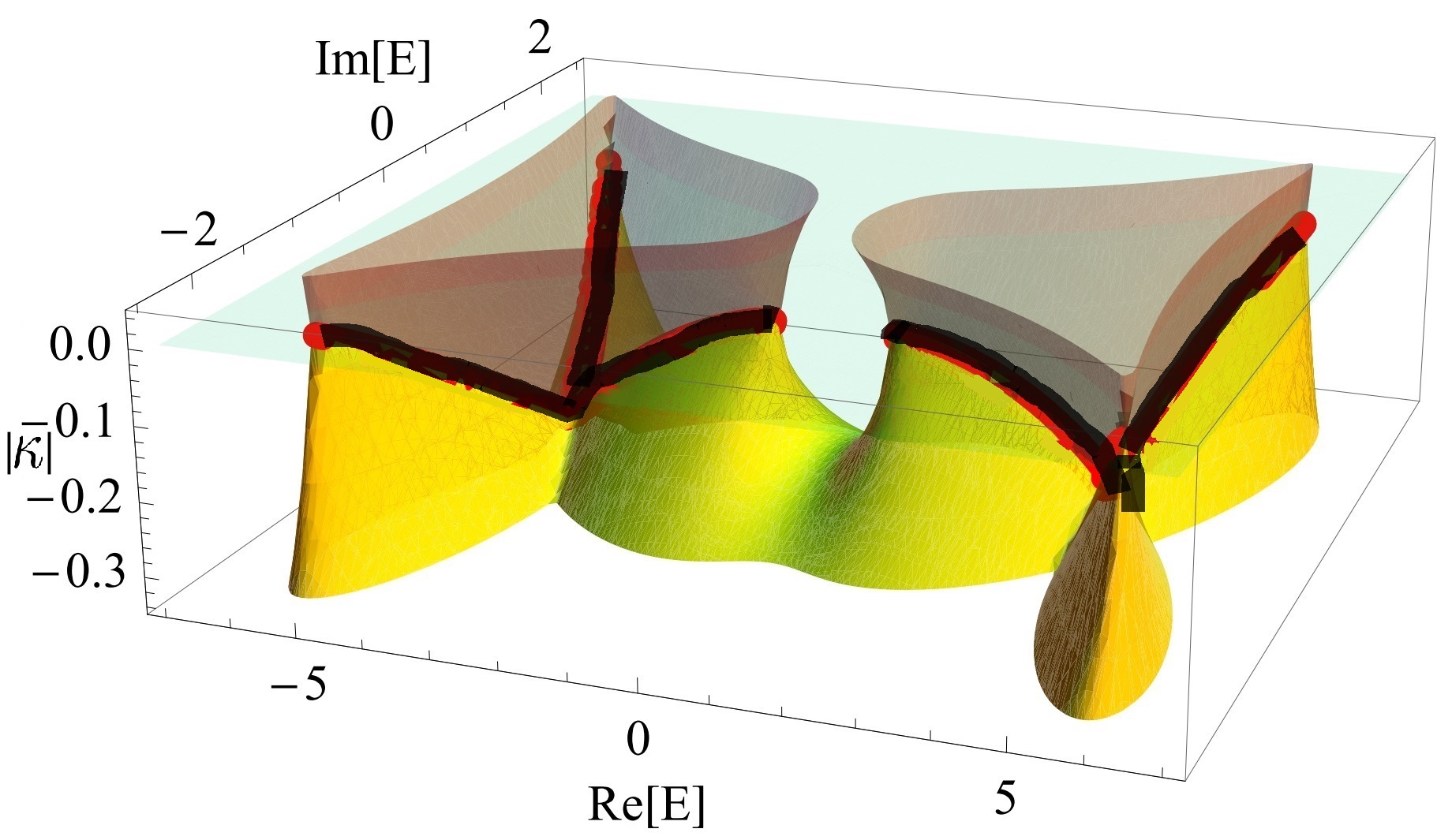}}\\
\end{minipage}
\caption{PBC-OBC spectral flow of model~\eqref{hmin}. In a), PBC bulk eigenvalues (red) flow along the blue-magenta curves, accumulating as OBC skin modes along the Y-shaped black lines. The latter are the largest magnitude solutions to $|\beta_\mu|=|\beta_\nu|$ (Eq.~\ref{beta}), as evident from the pale background contours at intervals $|\Delta\kappa|=\Delta \log|\beta|=0.07$. In b), this flow is visualized as eigenmodes "sliding down" the brown surface (solution of $|\beta|=e^{-\kappa}$) upon imaginary flux threading, till they stop along the black valley where two $|\beta|$ solutions (brown and yellow) intersect.
}
\label{fig:evolution}
\end{figure}
Necessary \emph{and} sufficient conditions for skin modes can be obtained by letting the imaginary flux $\kappa$ in Eq.~\ref{H_kappa} evolve from $0$ to $\pm \infty$, stopping when Eq.~\ref{beta} is satisfied for the first time. To appreciate the depth of this finding, we investigate a model with next-nearest neighbor (nnn) hopping, i.e., moving beyond Eq.~\ref{gamma_SSH}, from where the full complexity of non-Hermitian spectral flow unfolds: 
\begin{equation}
H_{\text{nnn}}(z)=\frac{9}{4}\sigma_x-3z\,\sigma_- +3\left(1-\frac1{z}-\frac1{z^2}\right)\sigma_+, \label{hmin}
\end{equation} 
$\sigma_\pm=(\sigma_x\pm i\sigma_y)/2$. 
From Fig.~\ref{fig:evolution}, its eigenmodes  (blue) flow towards the interior of the PBC energy loci (red), stopping only if they collide with other modes (black). Since these collisions occur at a single value of $\kappa$, they must be solutions where the $\beta$s with $e^{-\kappa}=|\beta|$ coincide (Eq.~\ref{beta}). Saliently, not all solutions of equal $|\beta|$ correspond to skin modes - only those with largest $|\beta|$, i.e., smallest $|\kappa|$ will be passed by the spectral evolution, and hence exist as OBC eigenmodes. All these observations hold for arbitrarily complicated Hamiltonians, reflecting the generic "contraction" property of imaginary flux flows~\cite{Suppmat}. In particular, models with PBC spectra already confined to lines or arcs, i.e., including all Hermitian and reciprocal systems (where $T_n=T^T_{-n}$), are precisely those without such flows, and hence skin effect.

\noindent{\it Criterion for non-Hermitian topological phases -- } 
Besides the continuum of skin boundary modes
, there can also exist isolated ``topologically protected'' boundary zero modes. The general criterion for their existence, however, must invariably differ in non-Hermitian systems from that of Hermitian models, since the skin effect introduces new decay length scales which manifest as additional singularities in the complex band structure. Below, we shall derive a novel topological criterion (Eq.~\ref{winding}) for the most intensely studied class of particle-hole (PH) symmetric 1D systems. 
It generalizes previously proposed invariants for non-Hermitian systems~\cite{yao2018edge,gong2018topological,yin2018geometrical,kunst2018biorthogonal}, and is straightforwardly applicable to models with arbitrarily complicated non-Hermitian hoppings.
Consider the most generic
PH symmetric 2-component Hamiltonian given by $H^{\text{PH}}[\{r_{a/b}\}; \{p_{a/b}\}](z)=$
\begin{equation}
\left(\begin{matrix}
0 & a(z) \\
b(z) & 0 \\
\end{matrix}\right)
=\left(\begin{matrix}
0 & z^{r_a}\prod_i^{p_a} \frac{(z-a_i)}{z\sqrt{a_i}} \\
z^{r_b}\prod_i^{p_b} \frac{(z-b_i)}{z\sqrt{a_i}}  & 0 \\
\end{matrix}\right),
\label{Ham}
\end{equation}
where $z=e^{ik}$ and $a_i,b_i$ 
are the complex roots of Laurent polynomials $a(z),b(z)$, both of which can be rescaled without changing the topology. In terms of OBC constraints (Eqs.~\ref{OBC1}), ``topological'' modes are special solutions where the boundary system described by Eqs.~\ref{OBC1} is not of full rank, such that 
the eigenmode weights $c_\mu$ have nonzero solutions \emph{despite} $|\beta_\mu|\neq |\beta_\nu|$ for any pair $\mu,\nu$. Rewriting Eqs.~\ref{OBC1} as a matrix equation $M\bf c=\bf 0$, this condition for a topological mode translates to $\text{Det}\,M=0$. As meticulously derived in the supplement~\cite{Suppmat}, this problem can be reformulated as the fundamental principle: {\it An isolated topological zero mode exists when the $r_a+r_b$ largest $\beta_\mu$s \emph{do not} contain $r_a$ members from $\{a_1,...,a_{p_a}\}$ and $r_b$ members from $\{b_1,...,b_{p_b}\}$.} 
These conditions on the zeros and poles of the Hamiltonian can also be recast~\cite{Suppmat} in terms of the windings 
\begin{align}
W_{g}(R)&=\oint_{|z|=R} \frac{d(\log g(z))}{2\pi i}=\#Z_g(R)-\#P_g,
\label{windings_ab}
\end{align}
$g=a,b$, which counts the number of zeros $\#Z_g(R)$ minus the number of poles $\#P_g$ encircled by a circle $|z|=R$ of radius $R\in\mathbb{R}$. Evidently, $\#P_g=p_g-r_g$ does not depend on $R$, since the poles are always at $z=0$. If $R$ is chosen such that $|z|=R$ excludes the $r_a$ largest roots of $a(z)$, $W_a(R_a)=(p_a-r_a)-\#P_a=0$ when a topological mode exists. The same $|z|=R$, however, is not allowed to simultaneously exclude $r_b$ roots of $b(z)$, for that would cause the $r_a+r_b$ excluded, i.e., largest roots, to be partitioned into $r_a$ $a_i$'s and $r_b$ $b_i$'s. Hence when $W_a(R)=0$, we must have $W_b(R)<0$, or vice versa. Thus a topological boundary mode exists iff
\begin{equation}
\exists\, R\in(0,\infty) \ \ \ \text{such that}\ \   W_a(R)W_b(R)<0,
\label{winding}
\end{equation}
or, in terms of the energy surface vorticity and eigenmode winding $V(R),W(R)=(W_a(R)\pm W_b(R))/2$~\cite{shen2018topological,yao2018edge},
\begin{equation}
\exists\, R\in(0,\infty) \ \ \ \text{such that}\ \  |V(R)|<|W(R)|.
\label{winding2}
\end{equation}
\begin{figure}[H]
\begin{minipage}{\linewidth}
\subfloat[]{\includegraphics[width=.56\linewidth]{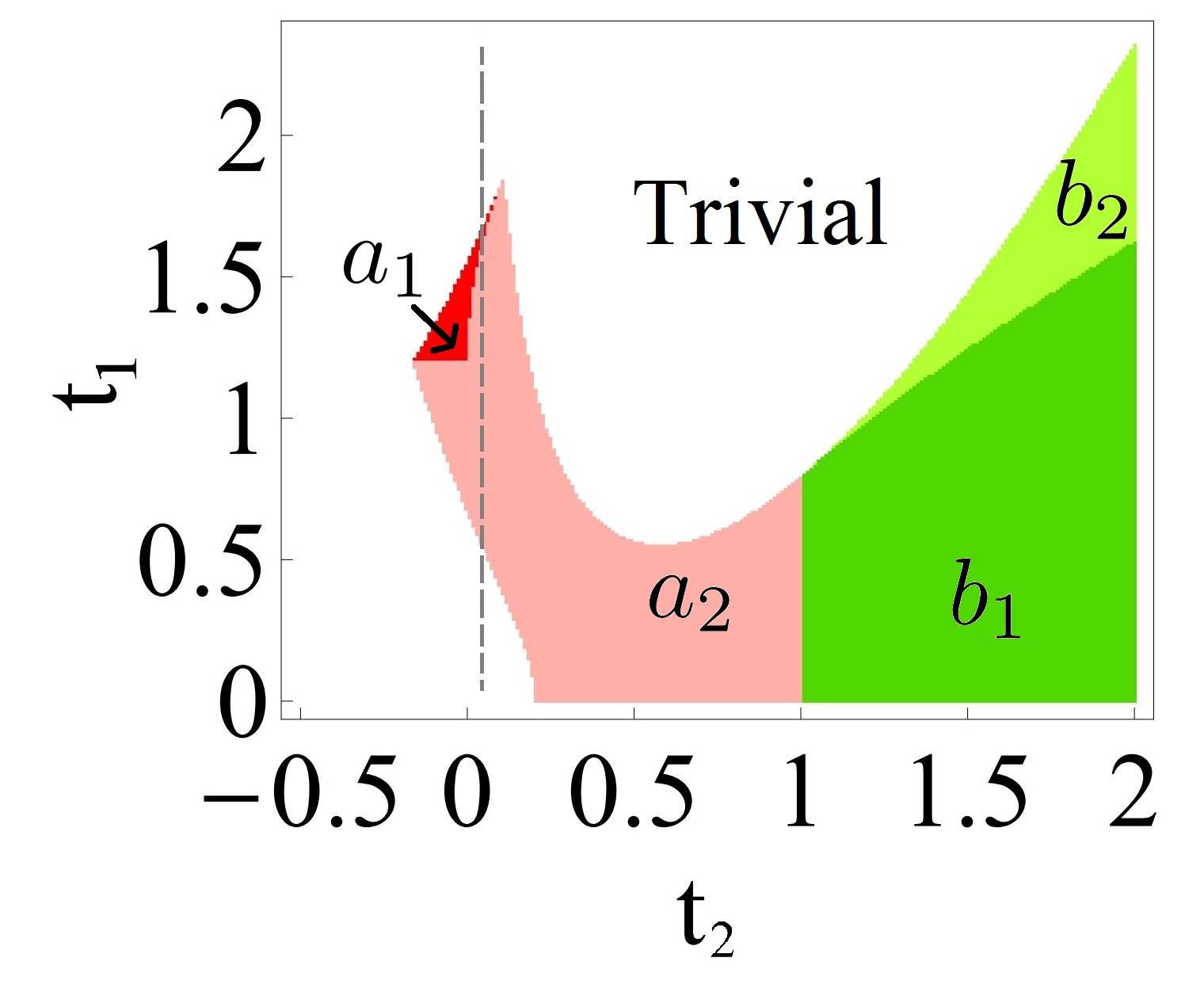}}
\subfloat[]{\includegraphics[width=.43\linewidth]{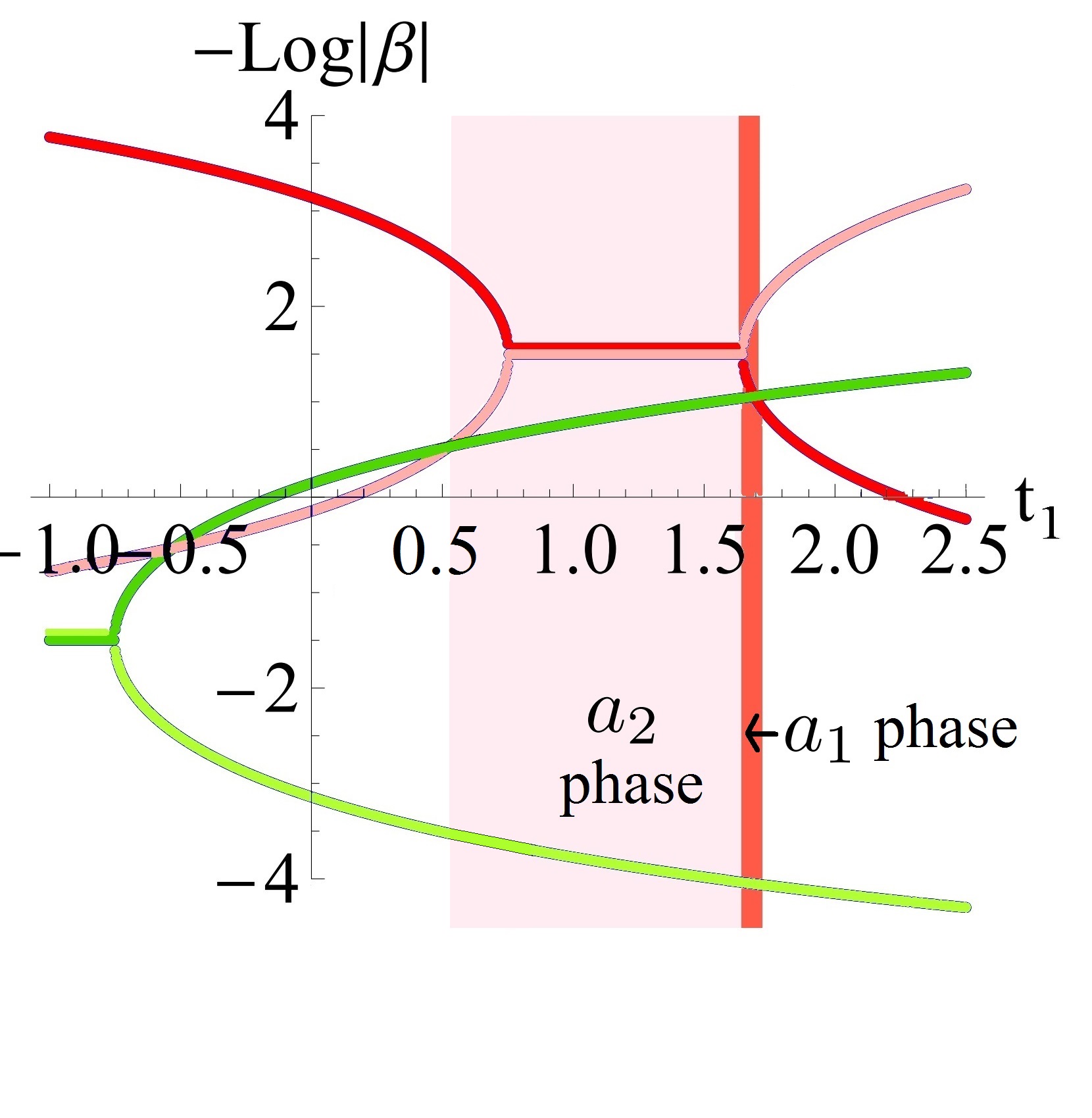}}
\end{minipage}
\caption{a) Phase diagram of Eq.~\ref{SSH3} with $\gamma=1.2$. Different colors represent regimes with topological mode decay rate $-(\log|\beta|)^{-1}$ determined by $\beta=a_1,a_2,b_1$ or $b_2$ respectively. b) Illustration of how the ordering of $-\log|\beta|$ solutions determine the phase along the dashed line ($t_2=0.05$) of a), with $\beta=a_1,a_2,b_1$ and $b_2$ solutions colored red, light red, dark green and light green. From criterion~\ref{winding}, topological modes occur when no greenish (redish) curve falls between two redish (greenish) curves, with corresponding regimes colored as in a).
}
\label{fig:topo3}
\end{figure}
\noindent

Criterion~\ref{winding} or~\ref{winding2} is a main result of this work, implying that to have topological modes, we need to find \emph{one} value of $R=e^{-\kappa}$ such that $W_a(R),W_b(R)$ are of opposite signs. Based on the insight that the OBC spectrum remains invariant under imaginary flux pumping, it does not rely on any specially tailored contour~\cite{yao2018edge}. As formulated in Eq.~\ref{winding2}, it expresses vorticity as a lower bound for eigenmode winding in the topological phase. For instance, when the energy surface contains a branch cut ($V(R)=1/2$), topological modes require the winding to be greater than $1/2$, not $0$ as in Hermitian cases.

To illustrate Eqs.~\ref{winding} and~\ref{winding2}, we apply it to a general nearest neighbor (nn) hopping model which is already beyond the models previously studied in the literature
\begin{equation}
H^{\text{PH}}_{\text{nn}}(z)=\left(\begin{matrix}
0 & t_1-\gamma + z + t_2/z\\
t_1+\gamma+1/z+t_2z & 0
\end{matrix}\right).
\label{SSH3}
\end{equation}
Its phase diagram (Fig.~\ref{fig:topo3}a) contains a topological region partitioned into four subregions, depending on whether the zero mode decay length $-(\log|\beta|)^{-1}$ is given by the roots $a_{1,2}=(-t_1-\gamma\pm \sqrt{(t_1+\gamma)^2-4t_2})/(2t_2)$, or $b_{1,2}=(\gamma-t_1\pm\sqrt{(t_1-\gamma)^2-4t_2})/2$. The decisive $\beta_\mu$ is the $(r_a+r_b+1)$th largest one~\cite{Suppmat} - not the one corresponding to the imaginary gap (largest $\beta_\mu$), which controls the hopping decays~\cite{lee2016band,lee2017band}, as illustrated in Fig.~\ref{fig:topo3}b. 
For $t_2=0$ in~\eqref{SSH3}, criterion~\ref{winding} reduces to previous formulations of a topological criterion~\cite{kunst2018biorthogonal,yao2018edge,jin2018bulk} $|t_1^2-\gamma^2|<1$ viz. $a_1=\infty,a_2=-\frac1{t_1+\gamma},b_1=\gamma-t_1$ and $b_2=0$. 

The fundamental advancement implied by criterion~\ref{winding} lies in its logical sufficiency, convenience of use and general applicability to all two-component PH-symmetric
\begin{figure}[H]
\includegraphics[width=.98\linewidth]{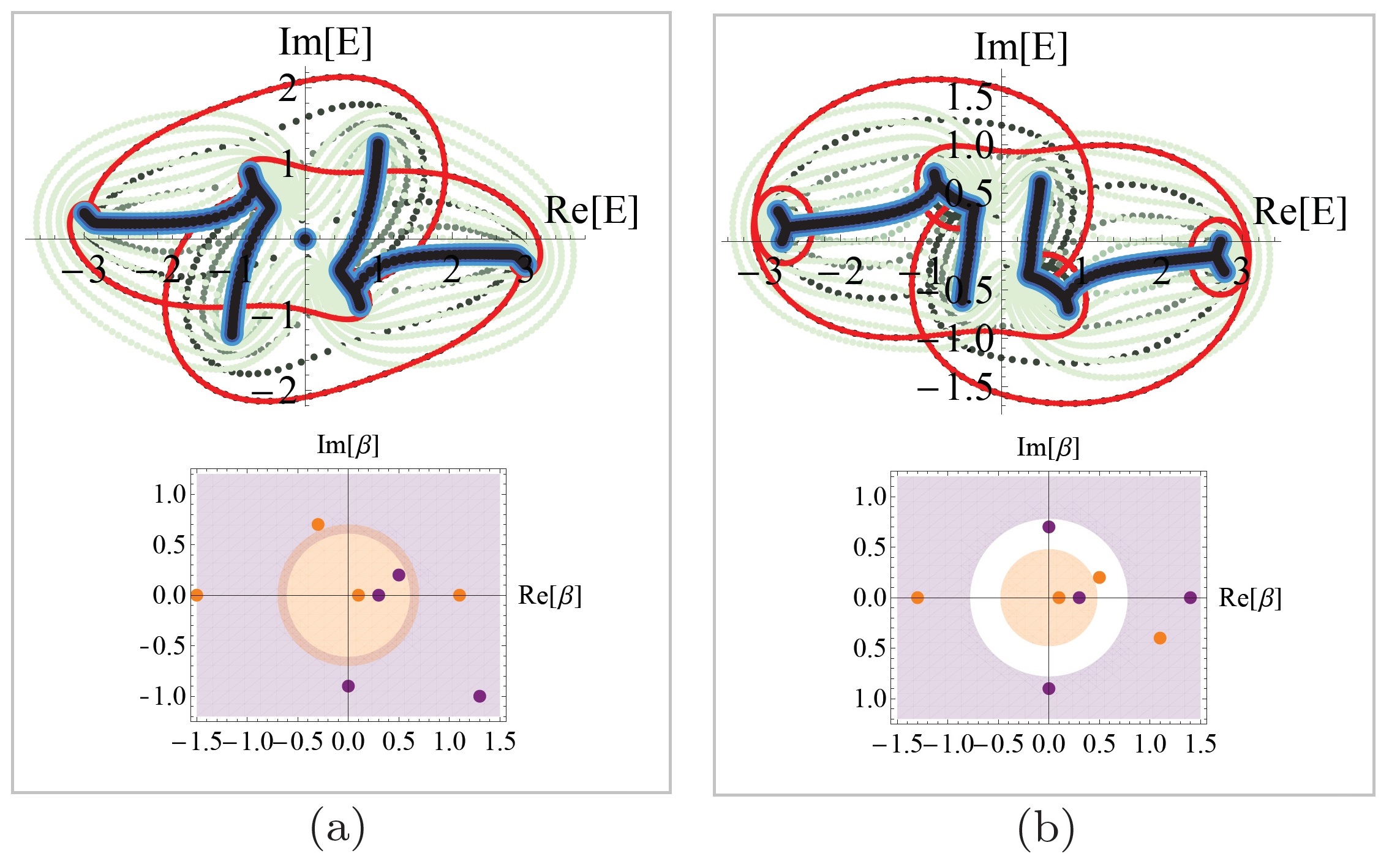}
\caption{Application of criterion~\ref{winding} to a more complicated instance of \eqref{Ham} with $p_a=p_b=4$ and $r_a=3$, $r_b=2$, which is completely topologically characterized by the roots of their $a(z)$ and $b(z)$ (purple and orange dots). 
Non-contractible contours in the purple region ($W_a>0$) enclose at least $q_a+1=p_a-r_a+1=2$ purple roots, while contours in the orange region ($W_b<0$) enclose fewer than $q_b+1=p_b-r_b+1=2$ orange roots. 
In a)/b), the presence/absence of a zero mode corresponds to the presence/absence of an overlap region where $W_a>0$ (purple) and $W_b<0$ (orange) simultaneously (i.e. $W_aW_b<0$).
}
\label{fig:topo4}
\end{figure}
\noindent Hamiltonians after finding the zeros. As demonstrated in Fig.~\ref{fig:topo4} for Hamiltonians with generic complex next-nearest neighbor hoppings and multiple roots, whether a zero mode exists depends on whether there exists a ring where $W_a>0$ and $W_b<0$ simultaneously (or vice-versa), i.e. where there are simultaneously less than $r_a$ larger zeros of $a(z)$ and less than $r_b$ smaller zeros of $b(z)$.   

\noindent{\it Discussion --} We have provided a rigorous treatment of boundary modes in non-Hermitian systems. We demonstrate how the skin modes can be characterized through an imaginary flux threading argument, and developed a winding number criterion for 1D topological boundary modes in PH-symmetric models. Our criterion $W_a(R)W_b(R)<0$ probes the \emph{entire} complex band structure, and, in the Hermitian case, reduces to the statement of nontrivial winding $W_a(R)^2>0$, where $W_b(R)=-W_a(R)$ and $R=1$. Our framework reveals the intuition behind the extreme sensitivity of non-Hermitian system to its boundary: even in a large system, a small reduction in the boundary hopping $\sim e^{-\kappa l}$ can be equivalent to a large change in $\kappa$ for the \emph{entire} system.

{\sl Acknowledgements --} We thank Zhong Wang, Xiao Zhang, Xiong Ye, Tobias Helbig and Tobias Hoffman for helpful comments. R.T. is supported by the European Research Council through ERC-StG-Thomale-TOPOLECTRICS-336012, by DFG-SFB 1170 (project B04), and by DFG-EXC 2471 "ct.qmat". 

\bibliographystyle{prsty}
\bibliography{references}

\clearpage

 \onecolumngrid
\begin{center}
\textbf{\large Supplemental Online Material for ``Anatomy of skin modes and topology in non-Hermitian systems" }\\[5pt]
Ching Hua Lee$^{1,2}$ and Ronny Thomale$^3$\\[5pt]
{\footnotesize \sl
$^1$Institute of High Performance Computing, A*STAR, Singapore, 138632.\\
$^2$Department of Physics, National University of Singapore, Singapore, 117542.\\
$^3$University of W\"urzburg, Am Hubland, D-97074 W\"urzburg, Germany
}
\end{center}
\vspace{0.1cm}
{\small This supplementary contains the following material arranged by sections:\\
\begin{enumerate}
\item Periodic-open boundary condition (PBC-OBC) evolution through imaginary flux - detailed derivations leading to key results Eqs.~1 and the discussion after Eq.~7 of the main text.
\item Pedagogical derivation of our topological criterion from first principles (Eqs.~11 and 12 of the main text).
\end{enumerate}
}
\setcounter{equation}{0}
\setcounter{figure}{0}
\setcounter{table}{0}
\setcounter{page}{1}
\setcounter{section}{0}
\makeatletter
\renewcommand{\theequation}{S\arabic{equation}}
\renewcommand{\thefigure}{S\arabic{figure}}
\renewcommand{\thesection}{S\Roman{section}}
\renewcommand{\thepage}{S\arabic{page}}
\vspace{0.2cm}

\section{PBC-OBC evolution through imaginary flux}

\subsection{Imaginary flux threading argument and semi-OBCs}

We treat a generic lattice system as a collection of 1D chains perpendicular to the open boundary, with coordinates of the other dimensions taken as external parameters. Consider a 1D chain described by a Hamiltonian 
\begin{equation}
H=\sum_{n=-N_L}^{N_R}\sum_{x;\gamma\delta} [T_n]_{\gamma\delta} \eta^\dagger_{x,\gamma}\eta_{x+n,\delta}=\sum_{n=-N_L}^{N_R}\sum_{k}\sum_{\gamma\delta}e^{ikn}[T_n]_{\gamma\delta}\eta^\dagger_{k,\gamma}\eta_{k,\delta},
\end{equation}
such that hoppings across a displacement of $n$ unit cells (i.e. sites) are given by the elements of the matrix $T_{n}$ in the sublattice (internal component) basis indexed by $\gamma,\delta$. $\eta^\dagger_{x,\gamma}$ and $\eta^\dagger_{k,\gamma}$ are the creation operators of a $\gamma$-th sublattice state at unit cell $x$ and quasi-momentum $k$ respectively. For brevity, we shall henceforth drop the sublattice indices. We assume reasonably local hoppings, so $N_L,N_R\sim O(1)$. 
Under periodic/open boundary conditions (PBCs/OBCs), the chain can be visualized as a ring with hoppings present/absent across its endpoints. Via Faraday's law, we can thread flux through this ring by shifting the momentum $k$ via minimal coupling $k\rightarrow k+\phi$, where $\dot\phi$ is the rate of change of flux which equals the induced (ficticious) electromagnetic field. Equivalently, this flux multiplies each hopping with a phase factor viz. $T_n\rightarrow T_ne^{in\phi}$. 

To relate this flux pumping with the boundary conditions (BCs), one performs a gauge transformation $H\rightarrow V^{-1}HV$ with $V=\text{diag}(e^{-i\phi},e^{-2i\phi},...,e^{-il\phi})$, $l$ being the system length. This removes the phase from all the hoppings except for those across the endpoints, which acquire a phase of $e^{\mp i l \phi}$. Through this, we have managed to re-express BCs on the boundary hoppings in terms of \emph{translationally-invariant} fluxes.

We next construct an interpolation between PBCs and OBCs for studying how non-Hermitian skin modes arise. For that, we have to first introduce the \emph{semi-open} boundary condition (semi-OBC), which has the boundary hoppings vanish in one direction but not the other. This is necessary because an imaginary flux component will always produce a rescaling factor $~O(e^{\pm l\,\text{Im}\,\phi})$ that diverges with $l$ at one of the boundaries. Without loss of generality, we set hoppings $T_{n<0}|_R$ from the right to the left boundary to zero, but preserve their reciprocal hoppings $T_{n>0}|_L$. As $\phi$ becomes complex, $T_{n>0}|_L$ will be rescaled by a factor of $e^{-l\,\text{Im}\,\phi}$. When $\text{Im}\,\phi=0$, we have perfect PBC in one direction; as $\text{Im}\,\phi\rightarrow \infty$, we approach the OBC limit. Had the non-reciprocity be directed in the opposite direction, an identical arguments holds with left and right sides switched, and $\phi\leftrightarrow \phi^*$.

Hence, to find the spectrum of $H(k)$ under the semi-OBC of $T_{n>0}|_R=0$ and $T_{n>0}|_L$ rescaled by a factor $e^{-\kappa l}$, which tends to the exact OBC when $\kappa l \rightarrow \infty$, we can perform the analytic continuation $k\rightarrow k+i\kappa$. In other words, we can simply diagonalize the \emph{translationally invariant} analytic continuation of the original Hamiltonian (Eq.~1 of the main text):
\begin{equation}
H_\kappa(k)=H(k+i\kappa),
\label{H_kappa}
\end{equation}
which possesses an identical spectrum as the semi-OBC system.  
Physically, \eqref{H_kappa} implies that 
all the original PBC bulk states \emph{must} morph into left boundary modes with localization lengths $\kappa^{-1}$ under $e^{-\kappa l}$ boundary hopping suppression. Furthermore, $H_\kappa(k)$ $\forall\,\kappa$ forms an equivalence class of Hamiltonians with identical OBC spectra. Such macroscopic condensation of modes onto one edge does not happen in Hermitian systems because semi-OBCs, being non-reciprocal, destroy hermiticity, and as such is a physically unrealistic proxy for OBC. But for the skin modes, OBCs and (correctly chosen) semi-OBCs are essentially equivalent, since the BCs are only consequential at the boundary where the skin mode is localized. Henceforth, we shall no longer distinguish OBCs from semi-OBCs.

\subsection{Geometric argument for when skin mode evolution stops (Eq.~7 and subsequent arguments of the main text)}

To intuitively understand why the eigenmodes should converge along exceptional points or arcs in the OBC limit, we consider their spectral flow upon threading of the \emph{real} part of a flux: $\phi=\text{Re}\, \phi+i\,\text{Im}\, \phi\rightarrow(\text{Re}\,\phi+2\pi/l)+i\,\text{Im}\, \phi $. 
This corresponds to multiplying the boundary hopping by a suppression factor together with a phase: $e^{-l\text{Im}\, \phi} \rightarrow e^{-l\text{Im}\, \phi }e^{2\pi i}$. Since $e^{2\pi i}=1$, this real flux evolution must map the full set of eigenvalues onto itself after a $2\pi$ period, i.e. it can only permute the eigenmodes. 

However, even this permutation should be trivial in the exact OBC limit of $\text{Im}\, \phi\rightarrow \infty$, since in this limit the boundary hopping disappears, and there will be no more boundary hopping to be rotated! As such, we intuitively expect the spectrum to contract into smaller and smaller loops when approaching the OBC limit (Fig.~\ref{fig:evolution}a), halting when the loops degenerate into arcs or isolated ``phenomenal'' exceptional points~\cite{xiong2018does,alvarez2018non} which exist only under OBCs and not PBCs. Hamiltonians which do not host skin modes are precisely those whose PBC spectra already are located along an arc. This includes all Hermitian systems, with spectra confined to the real line, as well as \emph{reciprocal} non-Hermitian systems, whose symmetric hoppings ($T_{n}=T^{T}_{-n}$) force the PBC spectrum to retrace itself. (Note that up to now, those are the models that have predominantely been realized in experimental setups.)

\newpage
\subsection{Examples}
Here we present more detailed results on the non-reciprocal SSH model ($\gamma_x=0,\gamma_y\neq 0$ from Eq. 2 of the main text). For convenience, we have defined $z=e^{ik}$ and $\sigma_\pm=(\sigma_x\pm i \sigma_y)/2$:
\begin{equation}
H_{SSH}^{\gamma_y}=\left(\frac1{2}+\cos k\right)\sigma_x +\sin k\,\sigma_y+i\gamma_y \sigma_y=\left(\frac1{2}+z^{-1}+\gamma_y\right)\sigma_++\left(\frac1{2}+z-\gamma_y\right)\sigma_-.
\label{SSHy}
\end{equation}
\begin{figure}[H]
\centering
\begin{minipage}{.99\linewidth}
\includegraphics[width=\linewidth]{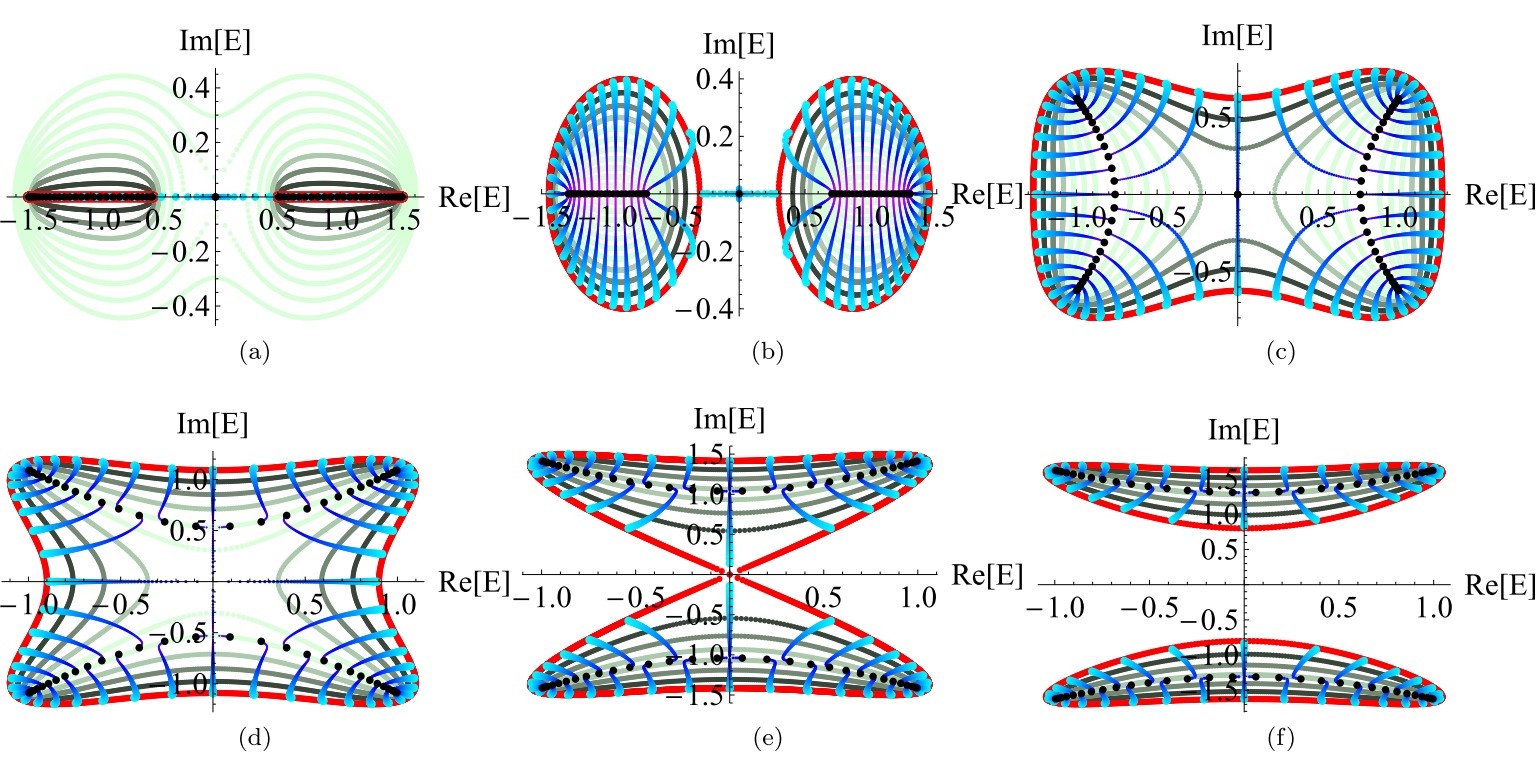}
\end{minipage}
\caption{(a-f) The $H_{SSH}^{\gamma_y}$ (Eq.~\ref{SSHy}) spectrum for $\gamma_y=0,0.4,0.8,1.2,1.5$ and $1.7$. As in the main text, the red curve represents the PBC spectrum and the blue/magenta tapering lines represent the PBC-OBC evolution trajectories of increasing $\kappa$, which collides to form the OBC spectrum (black). The pale background curves are contours of constant $\kappa$ with intervals of $0.1$. When $\gamma_y=0$ (a), we have the Hermitian SSH model, whose PBC and OBC spectra coincide except for the zero mode (black dot at $E=0$). As $\gamma_y$ increases, the PBC bands (red) broaden into ellipses (b). Before $\gamma_y$ exceeds $0.5$, the OBC limit can still be gauge transformed into that of the Hermitian SSH model\cite{yao2018edge}, and its spectrum (black) is hence confined to the real line. Beyond $\gamma_y=0.5$, the OBC spectrum also extends into the imaginary direction (c), finally annihilating with the zero mode and reopening in the perpendicular direction (d). This OBC topological phase transition occurs when the PBC spectrum is merged as a single loop (has nontrivial vorticity, i.e. is $4\pi$-periodic). Finally, at $\gamma_y=1.5$, the PBC gap also closes (e) before re-opening as the trivial phase (f). }
\label{fig:SSH_gamma}
\end{figure}

Next, we further study a more complicated next nearest neighbor hopping model (Eq. 8 of the main text):
\begin{equation}
H_{\text{nnn}}(z)=\frac{9}{4}\sigma_x-3z\,\sigma_- +3\left(1-\frac1{z}-\frac1{z^2}\right)\sigma_+, 
\label{hmin}
\end{equation} 
as well as a possible extension with third-nearest unit cell hoppings:
\begin{equation}
\tilde H_{\text{3rd nn}}(z)=\frac{9}{4}\sigma_x+\left(\frac{4}{z^3}-3z\right)\sigma_- +3\left(1-\frac1{z}-\frac1{z^2}\right)\sigma_+, 
\label{hmin2}
\end{equation} 
In these models, the higher powers of $z$ enable more complicated twists and turns in the PBC loop, although their OBC spectrum generally consist of relative straight sections (Fig.~\ref{fig:evolution}).

\begin{figure}[H]
\centering
\begin{minipage}{\linewidth}
\subfloat[]{\includegraphics[width=.48\linewidth]{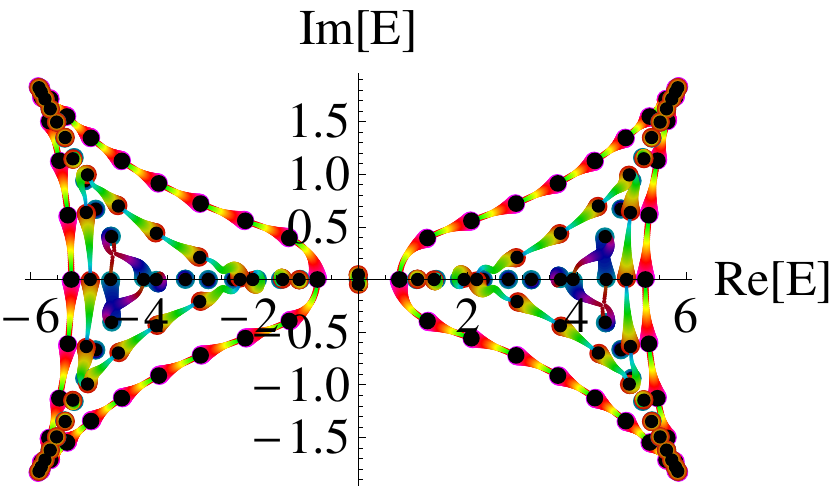}}\qquad
\subfloat[]{\includegraphics[width=.47\linewidth]{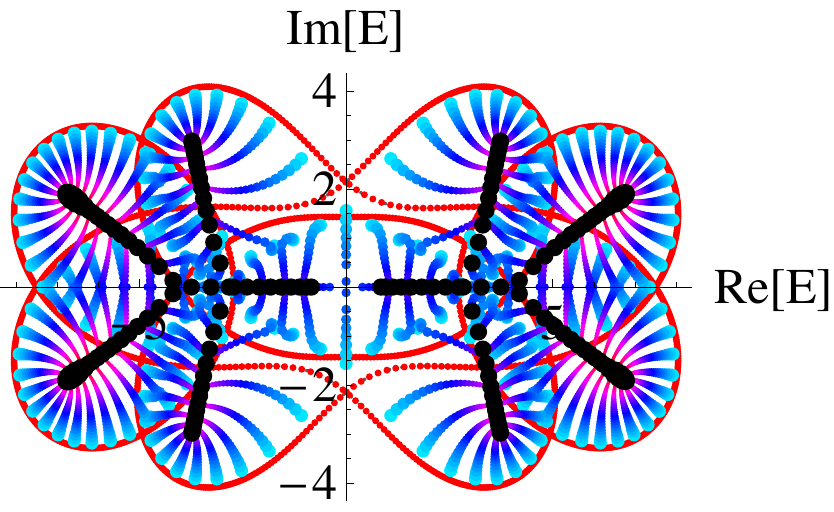}}
\end{minipage}
\caption{ a) Illustration eigenmodes of $H_{\text{min}}$ flowing into each other under the threading of a real flux $\text{Re}\,\phi\rightarrow \text{Re}\,\phi+2\pi/l$ (see above geometric argument), tracing successively smaller loops as the boundary hopping $~e^{-l\kappa}$ diminishes with increasing $\kappa$ (Shown are $\kappa=0,0.16,0.25$). For sufficiently large $\kappa$, each eigenvalue will flow into itself. b) PBC (red), OBC (black) and their interpolation trajectories (blue/magenta) for the extended minimal model $\tilde H_{\text{min}}$, which exhibit more convoluted loops which interpolate into more OBC branches hinged on by OBC exceptional points.}
\label{fig:evolution}
\end{figure}

\section{II. Derivation of the criterion for non-Hermitian particle-hole(PH) symmetric topological zero mode}
\subsection{General treatment of the open boundary condition}

\noindent In this section, we detail, from first principles, the detailed derivation of the topological criterion of particle-hole symmetric topological modes given by Eqs.~11 and 12 of the main text, as well as an equivalent formulation in terms of poles and zeros of the Hamiltonian. For a generic N-component Hamiltonian $H(z)$ where $z=e^{ik}$, any open boundary condition (OBC) eigenmode $\psi$ at eigenenergy $E$ can always be expanded in the Hilbert subspace of modes $\varphi_\mu$ that satisfy
\begin{equation}
H(\beta_\mu)\varphi_\mu=E\varphi_\mu,
\end{equation}
where $\beta_\mu$ is a root of the characteristic polynomial $\text{Det}[H(\beta)-E\,\mathbb{I}]=0$. Fourier transforming into real space, the OBC eigenmode $\psi$ can be written as
\begin{equation}
\psi(x)=\sum_\mu c_\mu \beta_\mu^x\varphi_\mu
\end{equation}
where the coefficients $c_\mu$ are chosen such that $H\psi(x)$ satisfies the OBC condition, i.e. vanishes outside an interval $x\in[0,l]$, where $l$ is the system length. Although there can be many more $\mu$'s than the number of bands in $H$, the basis spanned by $\beta_\mu^x\varphi_\mu$ is not necessarily overcomplete: This is because each $\beta_\mu^x\varphi_\mu$ for each different $x$ should be taken as a different basis mode. Although this sounds like an additional stringent requirement on $\psi(x)$, the OBC allows for certain spatially decaying solution modes that are prohibited under periodic boundary conditions (PBCs). Suppose that $H(z)$ contains hoppings of up to $N_L$ unit cells to the left, and up to $N_R$ unit cells to the right, i.e. 
\begin{equation}
H(z)=\sum_{-N_L<n<N_R}z^nT_n,
\end{equation}
where $T_n$ is the $N\times N$ hopping matrix across a displacement of $n$ sites (unit cells). The OBC will then constrain $\psi(x)$ for the $N_L$($N_R$) unit cells closest to the left(right) boundary. Specifically, terms that involve translations beyond the region $[0,l]$ must vanish. This yields Eq. 6 of the main text, which can be put into matrix form as $M\bf c=\bf 0$, where 
\begin{equation}
M=\left(\begin{matrix}
\sum_{1\leq n\leq N_L}\beta_1^{1-n}T_{-n}\varphi_1\,\, & \,\, \sum_{1\leq n\leq N_L}\beta_2^{1-n}T_{-n}\varphi_2 \,\, & \,\,  ...  \,\, & \,\,\sum_{1\leq n\leq N_L}\beta_{\mu_{max}}^{1-n}T_{-n}\varphi_{\mu_{max}}\\
\sum_{2\leq n\leq N_L}\beta_1^{2-n}T_{-n}\varphi_{\mu_1}\,\, & \,\, \sum_{2\leq n\leq N_L}\beta_2^{2-n}T_{-n}\varphi_{\mu_2} \,\, & \,\,  ...  \,\, & \,\,\sum_{2\leq n\leq N_L}\beta_{\mu_{max}}^{2-n}T_{-n}\varphi_{\mu_{max}}\\
\vdots\,\, & \,\, \vdots \,\, & \,\,  ...  \,\, & \,\,\vdots\\
(T_{1-N_L}+\beta_1^{-1}T_{-N_L})\varphi_1\,\, & \,\, (T_{1-N_L}+\beta_2^{-1}T_{-N_L})\varphi_2\,\, & \,\, ... \,\, & (T_{1-N_L}+\beta_{\mu_{max}}^{-1}T_{-N_L})\varphi_{\mu_{max}}\\
T_{-N_L}\varphi_1\,\, & \,\, T_{-N_L}\varphi_2\,\, & \,\, ... \,\, & T_{-N_L}\varphi_{\mu_{max}}\\
\sum_{1\leq n\leq N_R}\beta_1^{l+n-1}T_{n}\varphi_1\,\, & \,\, \sum_{1\leq n\leq N_R}\beta_2^{l+n-1}T_{n}\varphi_2 \,\, & \,\,  ...  \,\, & \,\,\sum_{1\leq n\leq N_R}\beta_{\mu_{max}}^{l+n-1}T_{n}\varphi_{\mu_{max}}\\
\sum_{2\leq n\leq N_R}\beta_1^{l+n-2}T_{n}\varphi_{\mu_1}\,\, & \,\, \sum_{2\leq n\leq N_R}\beta_2^{l+n-2}T_{n}\varphi_{\mu_2} \,\, & \,\,  ...  \,\, & \,\,\sum_{2\leq n\leq N_R}\beta_{\mu_{max}}^{l+n-2}T_{n}\varphi_{\mu_{max}}\\
\vdots\,\, & \,\, \vdots \,\, & \,\,  ...  \,\, & \,\,\vdots\\
(T_{N_R-1}+\beta_1^{l+1}T_{N_R})\varphi_1\,\, & \,\, (T_{N_R-1}+\beta_2^{l+1}T_{N_R})\varphi_2\,\, & \,\, ... \,\, & (T_{N_R-1}+\beta_{\mu_{max}}^{l+1}T_{N_R})\varphi_{\mu_{max}}\\
\beta_1^lT_{N_R}\varphi_1\,\, & \,\, \beta_2^lT_{N_R}\varphi_2\,\, & \,\, ... \,\, & \beta^l_{\mu_{max}}T_{N_R}\varphi_{\mu_{max}}\\
\end{matrix}\right)
\label{M}
\end{equation}
Here $M$ is an $\mu_{max}\times \mu_{max}$ matrix  where $\mu_{max}=N(N_R+N_L)$, since each entry above is an $N\times 1$ column vector. There are $\mu_{max}=N(N_R+N_L)$ unknown coefficients $c_\mu$, each corresponding to an eigenvector $\varphi_\mu$ and root $\beta_\mu$, since each entry of $H(\beta)$ is a Laurent polynomial in $\beta$ with up to $N_L+N_R$ roots, and each term in $\text{Det}[H(\beta)-E\,\mathbb{I}]$ is a product of $N$ such entries. Similarly, There are also up to $\mu_{max}=N(N_R+N_L)$ scalar constraints in the form of $N_R+N_L$ vector constraints, as explicated in the form of $M$ above. They consists of $N_L$ constraints from the sites near the left boundary (top half of $M$), and $N_R$ constraints from the sites near the right boundary (bottom half of $M$).
 
In general, a ``topological'' boundary mode corresponds to a solution to $M\bf c =\bf 0$ that is not part of an accumulation point set in the thermodynamic ($l\rightarrow \infty$) limit. In other words, it is an isolated solution that exists only if $\text{Det}M=0$. It is important to realize that, due to the finiteness of $l$, we must allow $E$ to be perturbed exponentially close (i.e. proportional to a power of $e^{-l}$) to its desired limiting value, which is $0$ when searching for zero modes. In a ``topological'' phase, it will be possible to find an exponentially small perturbation that satisfies $\text{Det}M=0$. Whether this perturbation exists depends on the locations of the roots of the characteristic polynomial, which can ultimately be cast in terms of so-called topological winding numbers. 

Although we shall explicitly treat only particle-hole symmetric 1D topological systems below, the above-mentioned relationship between topological winding numbers and boundary constraints is generally valid: Winding numbers encode bulk complex analytic properties which control what spatially decaying modes, which are necessary for satisfying boundary constraints, can exist. Since it is the asymptotic ($l\rightarrow \infty$) properties of these modes that play the pivotal role, details of the boundary constraints are largely immaterial. Hence the ``topological'' universality of these winding numbers.

\subsection{Specialization to PH symmetric 2-component non-Hermitian Hamiltonian}

To make the above treatment more concrete, we now specialize to finding zero modes ($E=0$ eigenenergies), and consider Hamiltonians of the form
\begin{equation}
H(z)=\left(\begin{matrix}
0 & a(z) \\
b(z) & 0 \\
\end{matrix}\right)
=\left(\begin{matrix}
0 & z^{-q_a}\prod_i^{p_a} (z-a_i) \\
z^{-q_b}\prod_i^{p_b} (z-b_i)  & 0 \\
\end{matrix}\right)
\end{equation}
where $z=e^{ik}$ and $a_i,b_i$ are the $p_a,p_b$ complex roots of $a(z),b(z)$ respectively. Here $N_L=\text{max}\{|q_a|,|q_b|\}$ and $N_R=\max\{r_a,r_b\}$, where $r_a=p_a-q_a$ and $r_b=p_b-q_b$. We have set the overall constants of $a(z)$ and $b(z)$ to unity, since their only effect is to rescale the energy trivially.

The particle-hole symmetry of $H(z)$ allows for considerable simplification of Eq.~\ref{M}. At $E=\pm\sqrt{a(z)b(z)}=0$, either $a(z)$ or $b(z)$ vanishes, and the $\mu_{max}=p_a+p_b$ roots of $\text{Det}\,H(\beta)=0$ are precisely the $a_i$ and $b_i$'s. Since $E=0$ is an exceptional point in this case, each $\beta_\mu$ correspond to only one normalized eigenvector $\varphi_\mu$, which must be of the form $(1,0)^T$ or $(0,1)^T$, depending on whether $\beta_\mu\in \{b_i\}$ or $\beta_\mu\in\{a_i\}$ respectively. Substituting these into Eq.~\ref{M} at $E\approx 0$ yields the following asymptotic form:
\begin{equation}
M\sim\left(\begin{matrix}
A_{1,1} & A_{1,2} & ... & A_{1,p_a} & 0 & 0 & ... & 0\\
0 & 0 & ... & 0 & B_{1,1} & B_{1,2} & ... & B_{1,p_b}\\
A_{2,1} & A_{2,2} & ... & A_{2,p_a} & 0 & 0 & ... & 0\\
0 & 0 & ... & 0 & B_{2,1} & B_{2,2} & ... & B_{2,p_b}\\
\vdots & \vdots & & \vdots & \vdots &\vdots && \vdots \\
A_{N_L,1} & A_{N_L,2} & ... & A_{N_L,p_a} & 0 & 0 & ... & 0\\
0 & 0 & ... & 0 & B_{N_L,1} & B_{N_L,2} & ... & B_{N_L,p_b}\\

A'_{1,1}a_1^l & A'_{1,2}a_2^l & ... & A'_{1,p_a}a_{p_a}^l & 0 & 0 & ... & 0\\
0 & 0 & ... & 0 & B'_{1,1}b_1^l & B'_{1,2}b_2^l & ... & B'_{1,p_b}b_{p_b}^l\\
A'_{2,1}a_1^l & A'_{2,2}a_2^l & ... & A'_{2,p_a}a_{p_a}^l & 0 & 0 & ... & 0\\
0 & 0 & ... & 0 & B'_{2,1}b_1^l & B'_{2,2}b_2^l & ... & B'_{2,p_b}b_{p_b}^l\\
\vdots & \vdots & & \vdots & \vdots &\vdots && \vdots \\
A_{N_R,1}a_1^l & A'_{N_R,2}a_2^l & ... & A'_{N_R,p_a}a_{p_a}^l & 0 & 0 & ... & 0\\
0 & 0 & ... & 0 & B'_{N_R,1}b_1^l & B'_{N_R,2}b_2^l & ... & B'_{N_R,p_b}b_{p_b}^l\\  
\end{matrix}\right)+O(E)
\label{M2}
\end{equation}
In the above, we have separated all the entries from Eq.~\ref{M} into constant scalars $A_{i,j},B_{i,j},A'_{i,j}$ and $B'_{i,j}$ which do not depend on the system size $l$, as well as factors $a_i^l$ and $b_i^l$ which decreases exponentially with $l$. In general, there are $q_a,q_b,r_a,r_b$ nonzero rows of the $A_{i,j},B_{i,j},A'_{i,j}$ and $B'_{i,j}$s respectively, adding up to $\mu_{max}=p_a+p_b$ constraints for $\mu_{max}$ unknown $c_\mu$ coefficients. The $O(E)=O\left(\sqrt{a(z)b(z)}|_{z\approx \beta_\mu}\right)$ correction arises from the small corrections from the $T_n\varphi_\mu$'s at $E$ slightly away from zero, which also decreases as a power of $e^{-l}$. 

\begin{figure}[H]
\begin{minipage}{\linewidth}
\subfloat[]{\includegraphics[width=.49\linewidth]{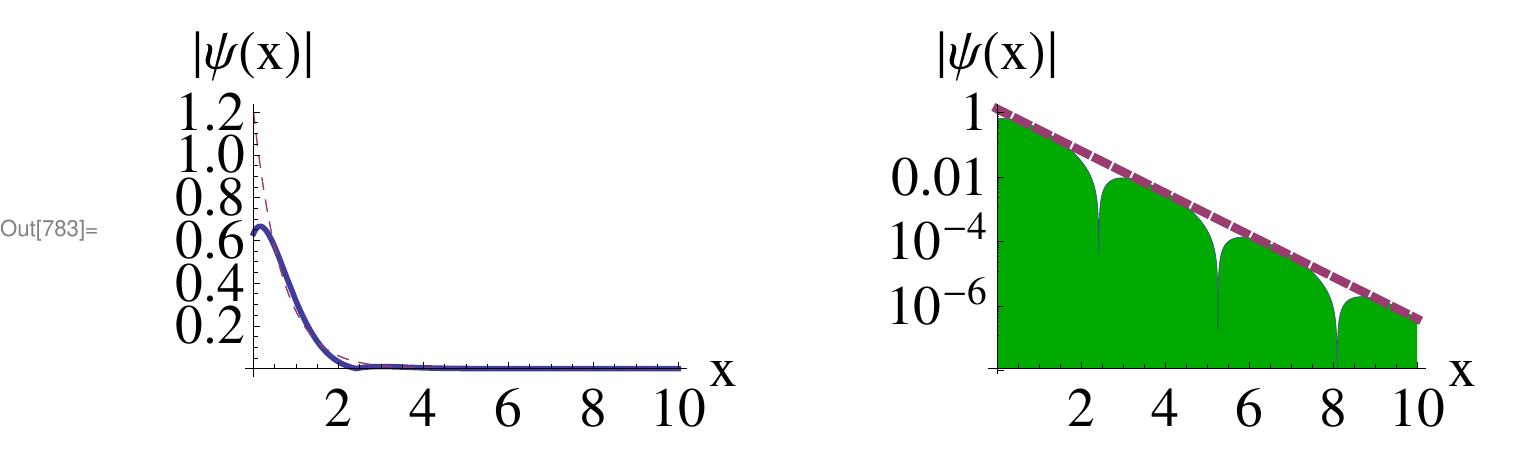}}
\subfloat[]{\includegraphics[width=.49\linewidth]{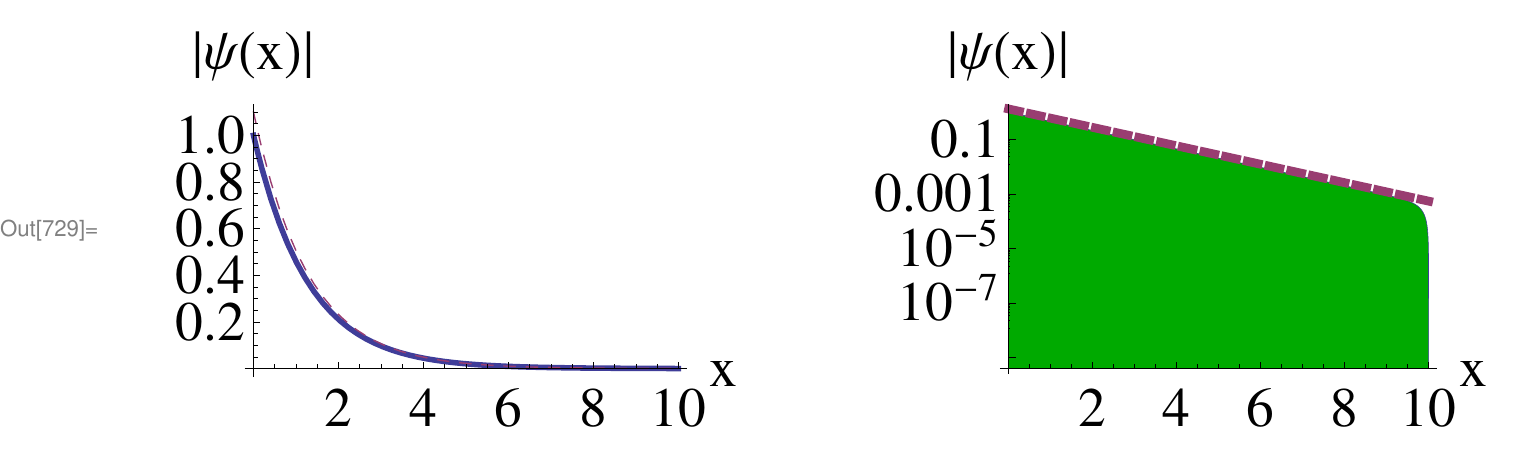}}
\end{minipage}
\caption{Two possible solutions to Eq.~\ref{M}, which approximately lie in the kernel of the matrix $M$. The Hamiltonian is given by Eq. 14 of the main text, with parameters $t_1=1,t_2=0.05$ and $\gamma=1.2$. 
}
\label{fig:decay}
\end{figure}

This separation of contributions with different scaling behaviors is the \emph{essential} step in the derivation of the topological criterion; other details of the entries of $M$ are inconsequential. From it, one can extract conditions on the $|a_i|,|b_i|$'s such that the edge mode condition $\text{Det}\,M=0$ is consistent with the scaling behavior; these conditions can then be recast in terms of winding numbers.

In the expansion of $\text{Det}\,M$, one necessarily have $l(r_a+r_b)$-degree monomials of the form 
\begin{equation}
\text{const.}\times\beta_1^l\beta_2^l...\beta_{r_a+r_b}^l,
\end{equation}
where $\{\beta_1,...,\beta_{r_a+r_b}\}\subset \{a_1,...,a_{p_a},b_1,...,b_{p_b}\}$. The monomials resulting from the expansion of the leading order matrix expression in Eq.~\ref{M2} necessarily contain $r_a$ of the $a_i^l$ s and $r_b$ of the $b_i^l$ s. However, the monomials from the $O(E^2)$ or higher order contributions can contain any number of the $a_i^l$ s and $b_i^l$ s, as long as there is a total of $r_a+r_b$ of them. For a boundary mode to exist, the roots $\beta_\mu$ need to be consistent with the fact that $E^2$ is exponentially decaying in $l$, while satisfying $\text{Det}\,M=0$. Below, we present two equivalent formulations for the criterion for satisfying the above requirements:

\subsection{Topological criterion: Decay length hierarchy formulation}

We order the roots $\beta_\mu\in \{a_1,...,a_{p_a},b_1,...,b_{p_b}\}$ of the characteristic polynomial $\text{Det}[H(\beta)-E\,\mathbb{I}]=0$ 
by \[|\beta_1|> |\beta_2|> |\beta_3| >...\]Physically, this is an ordering from the longest to shortest spatial decay length of their corresponding eigenmodes $\varphi_\mu$, which is given by $L_{\beta_\mu}=-(\log|\beta_\mu|)^{-1}$. Whether a boundary mode can exist or not depends entirely on the largest $r_a+r_b=p_a+p_b-q_a-q_b$ roots  $|\beta_1|,...,|\beta_{r_1+r_2}|$:
\begin{itemize}
\item For an isolated topological boundary mode to exist at $E=0$ when $l\rightarrow \infty$, we must \emph{not}, among the $r_a+r_b$ largest $\beta_\mu$'s, have $r_a$ of them belonging to $\{a_1,...,a_{p_a}\}$ and $r_b$ of them belonging to $\{b_1,...,b_{p_b}\}$\footnote{Equivalently, we can rephrase this as: Among the $q_a+q_b$ smallest $\beta_\mu$'s, we must \emph{not} have $q_a$ of them belonging to $\{a_1,...,a_{p_a}\}$ and $q_b$ of them belonging to $\{b_1,...,b_{p_b}\}$.}.
\end{itemize} 
If this criterion is violated, we will always find a monomial in the leading order contribution that contains all of the $r_a+r_b$ largest $\beta_\mu$'s. Since this is already the monomial with the largest possible magnitude, we will never be able to cancel it off with the subleading $O(E^2)$ contribution to give $\text{Det}\,M=0$ in the $l\rightarrow \infty$ limit.

Suppose that this criterion is respected. Let the largest roots be $\{a_1,...,a_{r_a+r_b-j}\}\bigcup\{b_1,...b_j\}$, where $j<r_b$. From Eq.~\ref{M2} and the arguments following it, the leading order monomial in the $O(E^0)$ contribution can only scale like $(a_1...a_{r_a})^l(b_1...b_{r_b})^l$. However, the $O(E^2)$ contribution generically contains every possible monomial, and will thus be dominated by $(a_1...a_{r_a+r_b-j})^l(b_1...b_{j})^l$. Hence $E^2$ will scale like
\begin{equation}
E^2\sim \frac{(a_1...a_{r_a})^l(b_1...b_{r_b})^l}{(a_1...a_{r_a+r_b-j})^l(b_1...b_{j})^l}=\left(\frac{\prod_{i=j+1}^{r_b}b_i}{\prod_{i=r_a+1}^{r_a+r_b-j}a_i}\right)^l
\label{E2}
\end{equation}
which always converges to zero. In terms of decay lengths $L_{a_i}$ and $L_{b_i}$'s, Eq.~\ref{E2} reads
\begin{equation}
\log |E|\sim -\frac{l}{2}\sum_{i=1}^{r_b-j}\left(\frac1{L_{b_{j+i}}}-\frac1{L_{a_{r_a+i}}}\right)\rightarrow -\infty
\end{equation}

Practically, we can directly obtain $|a_1|,...,|a_{p_a}|$ and $|b_1|,...,|b_{p_b}|$ from any given 2-band PH symmetric Hamiltonian. Of the $\binom{p_a+p_b}{p_a}$ possible partitions of these ordered roots into the two $a_i$ and $b_i$ sets, the above criterion gives
\begin{equation}
\sum^{p_a}_{j\neq q_a}\binom{q_a+q_b}{j}\binom{r_a+r_b}{p_a-j}=\binom{p_a+p_b}{p_a}-\binom{q_a+q_b}{q_a}\binom{r_a+r_b}{r_a}
\end{equation}
partitions that yield a boundary mode. Although all these partitions appears to give rise to the same topological zero mode, their decay lengths can differ. These results are also useful in the analysis of higher dimensional non-Hermitian systems like higher-order topological lattices~\cite{lee2018hybrid} and, more crucially, 3D non-Hermitian nodal metals~\cite{lee2018tidal}.

\subsection{Topological criterion: Winding number formulation}

The condition for the existence of isolated boundary modes $\text{Det}\, M=0$ will now be recast into an equivalent but more ``topological'' language (Eq.~\ref{winding2}). We define the winding numbers 
\begin{align}
W_a(R)&=\frac1{2\pi i}\oint_{|z|=R} d(\log a(z))\notag\\
W_b(R)&=\frac1{2\pi i}\oint_{|z|=R} d(\log b(z))
\label{windings}
\end{align}
which measure how many zeros minus the number of poles is encircled by each along the contour $|z|=R$. Evidently, the number of poles encircled by both do not depend on $R$, since $a(z)$ and $b(z)$ possess poles of order $q_a$ and $q_b$ at $z=0$ respectively. As for the zeros, we first choose an $R_a$ such that $|z|=R_a$ excludes the $r_a$ largest roots of $a(z)$. This gives $W_a(R_a)=(p_a-r_a)-q_a=0$. 
Now, in the previously formulated criterion for edge modes, the set of the largest $r_a+r_b$ roots of the characteristic polynomial \emph{cannot} be the union of the largest $r_a$ roots of $a(z)$ and the largest $r_b$ roots of $b(z)$; in other words, 
\begin{equation}
\text{min}\{|a_{r_a}|,|b_{r_b}|\}\notin \{|\beta_1|,...,|\beta_{r_a+r_b}|\}\,\,\,\, \Rightarrow \,\,\,\,\text{min}\{|a_{r_a}|,|b_{r_b}|\}< |\beta_{r_a+r_b}|,
\end{equation}
the above roots all ordered by magnitude. Hence $|z|=R_a$ must enclose at least one fewer root of $a(z)$ than of $b(z)$, or vice versa, i.e. if $W_a(R=R_a)=0$, $W_b(R=R_a)<0$, or vice versa. In a nutshell, 
\begin{itemize}
\item A topological boundary mode exists at $E=0$ when $l\rightarrow \infty$ iff
\begin{equation}
\exists\, R\in(0,\infty) \ \ \ \text{such that}\ \   W_a(R)W_b(R)<0
\label{winding2}
\end{equation}
where $W_a(R),W_b(R)$ are defined in Eq.~\ref{windings}. 
\end{itemize} 
Condition~\ref{winding2} is a new result that generalizes the topological criterion in Hermitian systems, where $W_b(R)=-W_a(R)$. In the latter, it reduces to the usual criterion of $W_a^2(R)>0$ for the existence of Hermitian topological modes, if one sets $R=1$. 

Eq.~\ref{winding2} can be expressed in terms of more familiar quantities~\cite{shen2018topological}:  The winding
\begin{equation}
W(R)=\frac{W_a(R)-W_b(R)}{2}
\end{equation} 
of the eigenmode $\propto \frac1{\sqrt{a(z)b(z)}}(b(z),a(z))^T$ as $z$ traces a circle of radius $R$ around the origin, and the vorticity
\begin{equation}
V(R)=\frac{W_a(R)+W_b(R)}{2}
\end{equation}
which gives the winding on the energy Riemann surface along the same contour; half-integer $V(R)$ signify a branch cut along a double-valued energy surface. It is trivial to show that $V^2(R)-W^2(R)=W_a(R)W_b(R)$, so that $W_a(R)W_b(R)<0$ is equivalent to $V^2(R)<W^2(R)$ or $|V(R)|<|W(R)|$ (Eq.~12 of the main text). In Hermitian systems, $V(R)$ always vanishes, and $|V(R)|<|W(R)|$ simply reduces to the usual condition of nonzero eigenmode winding.

\section{Detailed example: Non-Hermitian boundary mode from nearest-neighbor (NN) hoppings}

\noindent For pedagogical clarity, we provide the explicit mathematical details for a PH symmetric 2-component Hamiltonian of the form
 \begin{equation}
H(z)=\left(\begin{matrix}
0 & a(z) \\
b(z) & 0 \\
\end{matrix}\right)
\ =\ \left(\begin{matrix}
0 & (z-a_1)(z-a_2)/z\\
(z-b_1)(z-b_2)/z  & 0 \\
\end{matrix}\right)
\end{equation}
with $p_a=p_b=p=2$, $q_a=q_b=q=1$ and $\mu_{max}=p_a+p_b=4$ eigenmodes. Such Hamiltonians are simply enough to be analytically treated, but still possess sufficient richness for realizing most representive non-Hermitian phenomena. In more familiar notation, it is proportional to a generalized SSH model with complex coefficients:
\begin{eqnarray}
H^{\text{PH}}_{\text{min}}(z)=\left(\alpha_+\cos k +i\alpha_-\sin k-\alpha_0\right)\sigma_++[a_i\leftrightarrow b_i]\sigma_- 
\end{eqnarray}
where $\alpha_\pm=\sqrt{a_1a_2}\pm\frac1{\sqrt{a_1a_2}}$ and $\alpha_0=\sqrt{\frac{a_1}{a_2}+\frac{a_2}{a_1}}$ (remember that $a_1,a_2,b_1,b_2$ can all be complex). 

For any finite system size $l$, we expect the energy $E$ of a topological mode to be exponentially close to $0$, such that two solutions $\beta_1,\beta_2$ of the characteristic equation $\text{Det}[M-\mathbb{I}\,E]=0$ are approximately equation to the roots $a_1,a_2$ of $a(z)$. Likewise, $\beta_3\approx b_1$ and $\beta_4\approx b_2$. Their corresponding eigenmodes can be arbitrarily normalized since the $c_\mu$ coefficients can be rescaled at will, and we shall choose the following for convenience: 
\begin{equation}
\varphi_1=\left(\begin{matrix}
E \\ (a_1-b_1)(a_1-b_2)
\end{matrix}\right);\quad 
\varphi_2=\left(\begin{matrix}
E \\ (a_2-b_1)(a_2-b_2)
\end{matrix}\right);\quad
\varphi_3=\left(\begin{matrix}
(b_1-a_1)(b_1-a_2) \\ E
\end{matrix}\right);\quad 
\varphi_4=\left(\begin{matrix}
(b_2-a_1)(b_2-a_2) \\ E
\end{matrix}\right) \\
\end{equation}
Note that we have neglected the exponentially small differences between the $\beta_\mu$'s and the roots of $a(z)$ and $b(z)$, except when they are of leading order (as in $E$).
The translation hopping matrices from $H(z)$ are given by
\begin{equation}
T_1=\left(\begin{matrix}
0 & 1 \\
1 & 0 \\
\end{matrix}\right)\qquad \text{and}\qquad T_{-1}=\left(\begin{matrix}
0 & a_1a_2 \\
b_1b_2 & 0 \\
\end{matrix}\right).
\end{equation}
With them, we can construct the $M$ matrix representing the OBC constraints:
\begin{equation}
M=\left(\begin{matrix}
a_1a_2(a_1-b_1)(a_1-b_2) & a_1a_2(a_2-b_1)(a_2-b_2) & a_1a_2E & a_1a_2 E \\
b_1b_2E & b_1b_2E &  b_1b_2(b_1-a_1)(b_1-a_2) & b_1b_2(b_2-a_1)(b_2-a_2) \\
a_1^l(a_1-b_1)(a_1-b_2) & a_2^l(a_2-b_1)(a_2-b_2) & b_1^lE & b_2^l E \\
a_1^lE & a_2^lE &  b_1^l(b_1-a_1)(b_1-a_2) & b_2^l(b_2-a_1)(b_2-a_2) \\
\end{matrix}\right)
\end{equation}
A boundary mode can exist if $\text{Det}\,M=0$ can be satisfied. Explicitly, the latter is given by
\begin{eqnarray}
(a_1a_2b_1b_2)^{-1}\text{Det}\,M&=& -(a_1^l-a_2^l)(b_1^l-b_2^l)\left[\prod_{i,j=1}^2(a_i-b_j)^2+ E^4\right]\notag\\
&&+ E^2(a_1^la_2^l+b_1^lb_2^l)(a_1-a_2)(b_1-b_2)(a_1+a_2-b_1-b_2)^2\notag\\
&&+ E^2(a_1^lb_1^l+a_2^lb_2^l)(a_1-b_2)(a_2-b_1)((a_1-a_2)^2+(a_2-b_2)^2)\notag\\
&&- E^2(a_1^lb_2^l+a_2^lb_1^l)(a_1-b_1)(a_2-b_2)((a_1-b_2)^2+(a_2-b_1)^2)
\label{DetM2}
\end{eqnarray}
which is the explicit form of Eq.~\ref{M2} with all the higher order terms in $E^2$ written down. We see that although the leading order term does not contain monomials of the forms $a_1^la_2^l$ and $b_1^lb_2^l$,  the subleading $E^2$ contributions contains all $\binom{4}{2}=6$ types of monomials of degree $2l$. 

In the $l\rightarrow \infty$ limit, only the monomials containing the largest two $\beta_\mu\in\{a_1,a_2,b_1,b_2\}$ will dominate. Suppose that $|a_1|>|a_2|$ are the two largest. Then, since $a_1^la_2^l$ is absent from the $O(E^0)$ term, we can rearrange the dominant terms to obtain
\begin{eqnarray}
\text{Det}\,M=0\qquad\Leftrightarrow \qquad E^2|_{|a_1|>|a_2|>|b_1|>|b_2|}&\sim &\frac{(a_2^{-l}-a_1^{-l})(b_1^l-b_2^l)\prod_{i,j=1}^2(a_i-b_j)^2}{(a_1-a_2)(b_1-b_2)(a_1+a_2-b_1-b_2)^2}\notag\\
&\sim& \left(\frac{b_1}{a_2}\right)^l\frac{(a_1-b_1)(a_1-b_2)(a_2-b_1)(a_2-b_2)}{(a_1-a_2)(b_1-b_2)(a_1+a_2-b_1-b_2)^2}
\label{DetM3}
\end{eqnarray}
Since $|b_1|<|a_2|$, this is consistent with the requirement that $E\sim \left(\sqrt{\frac{b_1}{a_2}}\right)^l\rightarrow 0$ in the thermodynamic limit. This is a special case of Eq.~\ref{E2} with $j=0$ and $r_a=r_b=1$, although we have also evaluated the coefficient of the exponent.

Note that, if we had chosen say $a_1,b_1$ to be the largest two roots, in violation with the boundary mode criterion, the dominant monomial $a_1^lb_1^l$ would have appeared in the contributions at all orders of $E^2$, and $E^2$ will have to tend towards a finite value instead, i.e. not lead to a zero mode.

In summary, the exceptional nature of the $E=0$ in-gap point turns out to be key in expressing $\text{Det}\,M=0$ as a constraint on winding numbers. Exactly at $E=0$, either $a(z)$ or $b(z)$ vanishes and the $p_a+p_b$ roots $\beta_\mu$ correspond to the $a_i$'s or $b_i$'s, with corresponding eigenmodes $(1,0)^T$ or $(0,1)^T$. For any finite system size $l$, however, the topological mode is displaced from zero by $E\sim e^{-l}$, and the eigenmodes will acquire $O(E)$ corrections.

For a solution $\text{Det}\,M=0$ to exist, both sides of Eq.~\ref{DetM2} must scale similarly with $l$. Since the RHS is already suppressed by $E^2$, the LHS cannot contain the most slowly decaying terms. Specifically, the pairs $a_1,a_2$ or $b_1,b_2$ \emph{must} be the two $\beta_\mu$'s with largest magnitude, since they are absent in the LHS but not the RHS. This leads to the result of Eq.~\ref{DetM3}. Such constraints imposed by the scaling suppression from $E^2$ also appear in generic cases, and is guaranteed by the defective eigenspace of $H(z)$ at $E=0$.

\subsubsection{Simplest case of the non-reciprocal SSH model}
In the special simplest case of the non-reciprocal SSH model (Eq.~\ref{SSHy}), 
\begin{equation}
H_{SSH}^{\gamma_y}(z)=\left(\begin{matrix}
0 & t+\gamma+\frac1{z} \\
t-\gamma+z & 0 \\
\end{matrix}\right)
\ = \ \left(\begin{matrix}
0 & (t+\gamma)\frac{(z+\frac1{t+\gamma})}{z}\\
\frac{(z-(\gamma-t))(z-0)}{z}  & 0 \\
\end{matrix}\right)
\end{equation}
and, after discarding inconsquential scalar factors, we identify $a_1=\infty$, $a_2=-\frac1{t+\gamma}$, $b_1=\gamma-t$ and $b_2=0$. The boundary mode criterion states that the two largest roots must be either $a_1,a_2$ or $b_1,b_2$. But since $a_1$ and $b_2$ are already fixed, the only option is to have let them be $a_1,a_2$. Hence $|a_2|>|b_1|$, i.e. we need 
\begin{equation}
\frac1{|t+\gamma|}> |\gamma-t|\qquad \Rightarrow \qquad |t^2-\gamma^2|<1
\end{equation} 
for a topological mode in the $\gamma-SSH$ model, in agreement with the literature\cite{yao2018edge,kunst2018biorthogonal}.

In terms of the equivalent winding criterion Eq.~\ref{winding2}, we have
\begin{gather}
W_a(R)=\begin{cases}
  0\qquad \text{if}\quad R>\frac1{|\gamma+t|}\notag\\    
  -1 \quad \text{if} \quad R<\frac1{|\gamma+t|}   
\end{cases}
\end{gather}
\begin{gather}
W_b(R)=\begin{cases}
    1\qquad \text{if}\quad R>|\gamma-t|\notag\\    
  0 \qquad \text{if} \quad R<|\gamma-t|   
\end{cases}
\end{gather}
To have $W_a(R)W_b(R)<0$, the $W_a(R)=-1$ region must overlap with the $W_b(R)=1$ region. This is possible if there exists $R$ such that $|\gamma-t|<R<\frac1{|\gamma+t|}$, i.e. the same conclusion $|t^2-\gamma^2|<1$.


\end{document}